\documentclass{iaus}
\usepackage{graphicx}

\title[Red halos of galaxies] %% give here short title %%
{Red halos of galaxies -- Reservoirs of baryonic dark matter?}

\author[Zackrisson et al.]   %% give here short author list %%
{E. Zackrisson$^{1, 2, 3}$, N. Bergvall$^3$, C. Flynn$^1$, \break G. \"Ostlin$^2$, G. Micheva$^2$ \and B. Caldwell$^3$}

\affiliation{$^1$Tuorla Observatory, University of Turku, V\"ais\"al\"antie 20, FI-21500 Piikki\"o, Finland\\[\affilskip]
$^2$Stockholm Observatory, AlbaNova University Center, 106 91 Stockholm, Sweden\\[\affilskip]
$^3$Department of Astronomy and Space Physics, Box 515, 751 20 Uppsala, Sweden\\[\affilskip]}

\pubyear{2007}
\volume{244}  %% insert here IAU Symposium No.
\pagerange{1--9}
\date{}
\jname{Dark Galaxies and Lost Baryons}
\editors{XXX, eds.}
\begin{document}

\maketitle

\begin{abstract}
Deep optical/near-IR surface photometry of galaxies outside the Local Group have revealed faint and very red halos around objects as diverse as disk galaxies and starbursting dwarf galaxies. The colours of these structures are too extreme to be reconciled with stellar populations similar to those seen in the stellar halos of the Milky Way or M31, and alternative explanations like dust reddening, high metallicities or nebular emission are also disfavoured. A stellar population obeying an extremely bottom-heavy initial mass function (IMF), is on the other hand consistent with all available data. Because of its high mass-to-light ratio, such a population would effectively behave as baryonic dark matter and could account for some of the baryons still missing in the low-redshift Universe. Here, we give an overview of current red halo detections, alternative explanations for the origin of the red colours and ongoing searches for red halos around types of galaxies for which this phenomenon has not yet been reported. A number of potential tests of the bottom-heavy IMF hypothesis are also discussed.
\end{abstract}

\firstsection % if your document starts with a section,
\section{Introduction}
The quest to unravel the nature of the dark matter, estimated to account for $\approx 90\%$ of the matter content of the Universe (e.g. \cite[Spergel et al. 2007]{Spergel et al.}), remains one of the most important tasks of modern cosmology. The dark matter appears to exist in at least two separate forms: one baryonic, and one non-baryonic. While the non-baryonic component is the dominant one, a substantial fraction of the baryons ($\approx 1/3$; \cite[Fukugita 2004]{Fukugita}; \cite[Fukugita \& Peebles 2004]{Fukugita & Peebles}) are still at large in the local Universe. 

The old idea of baryonic dark matter in the form of faint, low-mass stars has recently gained new momentum through surface photometry detections of very red and exceedingly faint structures -- so-called red halos -- around different types of galaxies. The history of this topic goes back to the mid-90s, when deep optical and near-IR images (e.g. \cite[Sackett, Morrison, Harding, \etal\ 1994]{Sackett et al.}; \cite[Lequeux, Fort, Dantel-Fort, \etal\ 1996]{Lequeux et al.}; \cite[Rudy, Woodward, Hodge, \etal\ 1997]{Rudy et al.}; \cite[James \& Casali 1998]{James & Casali}) indicated the presence of a faint halo around the edge-on disk galaxy NGC 5907. The colours of this structure were much too red to be reconciled with any normal type of stellar population, and indicative of a halo population consisting primarily of low-mass stars. At around the same time, \cite{Molinari et al.} also announced the detection of a red halo around the cD galaxy at the centre of Abell 3284. Skepticism grew with the discovery of what appeared to be the remnants of a disrupted dwarf galaxy close to NGC 5907 (\cite[Shang, Brinks, Zheng, \etal\ 1998]{Shang et al.}) and suggestions that this feature, in combination with other effects, could have resulted in a spurious halo detection (\cite[Zheng, Shang, Su, \etal\ 1999]{Zheng et al.}). While follow-up observations with the Hubble Space Telescope (HST) took some of the edge out of this criticism (\cite[Zepf, Liu, Marleau, \etal\ 2000]{Zepf et al.}), the red halo of NGC 5907 remains controversial. \cite{Yost et al.} failed to detect any red halo at wavelengths of 3--5$\mu$m, but since their upper limit of 15\% on the contribution from hydrogen-burning stars to the overall dark matter of this galaxy is more or less identical to the cosmic baryon fraction ($\Omega_\mathrm{baryons}/\Omega_\mathrm{M}\approx 0.15$; \cite[Spergel, Bean, Dor\'e, \etal\ 2007]{Spergel et al.}), this constraint is no longer relevant for a red halo of low-mass stars contributing to the {\it baryonic} component of the dark matter.

A few years later, \cite{Bergvall & Östlin} and \cite{Bergvall et al.} discovered similar faint and abnormally red structures in deep optical/near-IR images of blue compact galaxies (BCGs). Even more recently, \cite{Zibetti et al.} stacked the images of 1047 edge-on disk galaxies from the Sloan Digital Sky Survey (SDSS) and detected a halo population with a strong red excess and optical colours curiously similar to those previously derived for NGC 5907 -- again very difficult to reconcile with standard stellar populations. The halo detected around an edge-on disk galaxy at redshift $z=0.322$ in the Hubble Ultra Deep Field shows similarly red colours (\cite[Zibetti \& Ferguson 2004]{Zibetti & Ferguson}).

% NOTE use of \upi in above paragraph and subsequently throughout paper.
% The Greek constant character pi should be upright.

\section{A bottom-heavy stellar initial mass function?}
\cite{Zackrisson et al.} analyzed the colours of these new detections and found that the halos of both BCGs and edge-on disks could be explained by a stellar population with a very bottom-heavy IMF ($dN/dM\propto M^{-\alpha}$ with $\alpha\approx 4.50$, where $\alpha=2.35$ represents the Salpeter slope). For an IMF slope as extreme as this, only stars with masses $0.1\leq M\ (M_\odot)\leq 3$ contribute substantially to the integrated optical/near-IR light. The mass-to-light ratios of such populations are very high ($M/L_B \gtrsim 40$), which make them potential reservoirs for some of the ``dark'' baryons still missing in the low-redshift Universe. Corroborating evidence for the viabilitiy of an IMF this bottom-heavy comes from star counts in the field population of the LMC, where a slope of $\alpha\approx 5$--6  has been derived for masses $\geq 1 \ M_\odot$ (\cite[Massey 2002]{Massey}; \cite[Gouliermis, Brandner \& Henning 2006]{Gouliermis et al.}). The characteristic mass of the putative red halo stars ($0.1\leq M\ (M_\odot)\leq 1$) also coincides with that of the claimed MACHO detections in the halos of the Milky Way and M31 (\cite[Alcock, Allsman, Alves, \etal\ 2000]{Alcock et al.}; \cite[Calchi Novati, Paulin-Henriksson, An, \etal\ 2005]{Calchi-Novati}), although the interpretation of these microlensing observations remain controversial (e.g. \cite[Tisserand, Le Guillou, Afonso, \etal\ 2007]{Tisserand et al.}). The overall mass of the red halo structure is unfortunately very difficult to assess from current data, since the red halo signal has so far only been detected over a limited spatial region of each galaxy. Even if the mass-to-light ratio of the red halo is assumed to be independent of radius, its surface brightness profile would need to be extrapolated inwards and outwards in order to derive the total mass.

\section{Ongoing searches}
To get to the heart of the red halo mystery, our group has begun a large number of observational searches for red halos around types of galaxies for which this phenomenon has not yet been reported. This includes post-starburst galaxies (PI Zackrisson), elliptical galaxies (PI Bergvall), and dwarf galaxies in the Local Group (PI Zackrisson). We are also obtaining multiband data for an extended sample of BCGs (PI \"Ostlin). If some classes of objects systematically turn out to have red halos, whereas others don't, this will provide important clues to the formation of these structures. A non-detection of red halos around post-starbursts and ellipticals would, for instance, indicate that the red halo phenomenon is related to active star formation, suggesting -- perhaps -- that the red excess of these halos is of interstellar, rather than stellar, origin.  

While \cite[Zibetti \etal\ (2004)]{Zibetti et al.} detected a red halo around edge-on disk galaxies through the stacking of {\it high} surface brightness disks the SDSS, our team has recently carried out the same procedure for $\approx 1500$ {\it low} surface brightness disks (\cite[Caldwell \& Bergvall 2007]{Caldwell \& Bergvall}; Bergvall \etal\, in preparation). This resulted in the detection of a red halo with colours even more extreme than the ones previously discovered around edge-on disks (both the stacked SDSS disks and NGC 5907). The fact that a red excess (compared to what can easily be explained by a stellar population with a Salpeter-like IMF) has now been seen in a completely independent sample of stacked disks suggests that the Zibetti \etal\ detection is unlikely to be a statistical fluke (i.e. a ``$2\sigma$ effect''). Whatever its origin, the red halo phenomenon seems to be here to stay. 
 
\section{Potential tests}
The hypothesis that the red halo colours are generated by a stellar population with an abnormally high fraction of low-mass stars can in principle be tested using a number of different methods, which we outline below.   

\begin{enumerate}
\item {\bf Improved optical/near-IR data}. Once red halo data in more filters (e.g. $BVRI$\\$JHK$) become available, the bottom-heavy IMF may no longer be able to explain the spectral energy distribution, and can then be rejected. With H$\alpha$ and H$\beta$ emission-line data (obtained through narrowband photometry), the contribution from nebular emission at large projected distances from the centre of the target galaxies can also be assessed. This test is currently being implemented using very deep multiband data (including H$\alpha$ and H$\beta$) obtained at the 2.5m Nordic Optical Telescope (PI Zackrisson) for UM456, a BCG for which previous, shallower data has already revealed the presence of a very red halo. Near-IR surface photometry data obtained from space (e.g. with AKARI) would also be very useful, as this could alleviate some of the difficulties associated with ground-based measurements of the red halo signal against the very bright near-IR sky.
\item {\bf Mid-IR data}. A large population of preferentially low-mass stars should be detectable at near/mid-IR wavelengths with the Spitzer Space Telescope. Spitzer data for the halo regions of a handful of BCGs has already been obtained (PI Bergvall), and the analysis is underway.   
\item {\bf Spectroscopy of red halos}. While extremely challenging at the faint surface brightness levels at which the red halos are detected, spectroscopy could potentially reveal signatures in the spectral energy distribution of the red halos that are inconsistent with a bottom-heavy IMF. Integral-field spectroscopy with VLT/VIMOS has recently been attempted for a handful of BCG halos (PI Bergvall).
\item {\bf Star counts in nearby galaxies}. Since a bottom-heavy IMF would generate a much smaller fraction of red giants compared to a normal (i.e. Salpeter-like) IMF, the bottom-heavy IMF hypothesis could be falsified through direct star counts in halo regions for which surface photometry data have already revealed extremely red colours. Individual giants can be detected out to distances of $\approx 10$ Mpc with the HST and very deep data is already available in the HST archive for a number of BCGs within this distance. The ground-based surface photometry required to perform this test has been carried out at the 2.5m Nordic Optical Telescope (PI Zackrisson). A test of this type has already been carried out by \cite[Zepf \etal\ (2000)]{Zepf et al.} for the red halo of NGC 5907, with the intriguing result that the small number of giants detected indeed seemed to favour a bottom-heavy IMF. Because of the controversial nature of the surface photometry data for this object, it is, however, not entirely clear how robust this conclusion really is. An absence of red giants would moreover not rule out a non-stellar explanation for the red colours observed.
\item {\bf Star counts in Local Group galaxies}
While the previous methods all have the potential to falsify a bottom-heavy IMF as the origin of the observed red halo colours, none of them will be able to provide a direct confirmation of the presence of unusually many low-mass stars. To accomplish this, one has to turn to objects within the Local Group, where stars at masses $< 1 \ M_\odot$ can be individually resolved. Unfortunately, no red halos have yet been discovered (or looked for) among the dwarf galaxies of the Local Group. The main reason for this is that deep surface photometry becomes exceedingly difficult for galaxies with large angular sizes. If Local Group dwarf galaxies do exhibit red halos, these are likely to completely fill the field of view of current detectors (about $10^\prime\times 10^\prime$ for typical CCDs and $30^\prime\times 30^\prime$ for wide field cameras), leaving few or no regions in the resulting image from which a clean measurement of the sky flux can be made. To be able to reliably subtract the sky down to the very faint surface brightness levels where red halos are expected to be seen, one would either have to employ very rapid sky chopping (even in the optical), or alternatively target only part of the halo region, hoping that any large-scale gradients in the sky flux can be corrected for. While admittedly challenging, we are currently attempting this at the ESO 2.2m/WFI (PI Zackrisson). If successful, follow-up star counts should be able to robustly confirm or reject the connection between low-mass stars and the red excess. Deep HST images are available for many of the Local Group dwarfs, but with a few exceptions, these have been obtained too close to the centres of these galaxies, where a red halo signal in unlikely to be detectable. 
\end{enumerate}

\section{A red halo of low-mass stars around the Milky Way?}
While the identification of a red halo around stacked high-surface brightness disk galaxies in the SDSS (\cite[Zibetti \etal\ 2004]{Zibetti et al.}) indicates that a substantial fraction of disks are accompanied by red halos, this does not necessarily mean that {\it all} disk galaxies have a red structure of this type. It is nonetheless interesting to ask whether the Milky Way itself could be surrounded by a hitherto undetected red halo of low-mass stars, with photometric properties similar to that detected around stacked edge-on disks. The detected baryonic components (thin disc, thick disk, bulge and known stellar halo) of the Milky Way contribute around 5--$6\times 10^{10}\ M_\odot$ (e.g. \cite[Sommer-Larsen \& Dolgov 2001]{Sommer-Larsen & Dolgov}; \cite[Klypin, Zhao \& Sommerville 2002]{Klypin et al. b}) to its virial mass of $\approx 1\times 10^{12} \ M_\odot$ (\cite[Klypin \etal\ 2002]{Klypin et al. b}). A cosmic baryon fraction of $\Omega_\mathrm{baryons}/\Omega_\mathrm{M}\approx 0.15$ (\cite[Spergel \etal\ 2007]{Spergel et al.}), combined with the theoretical prediction that the baryon fraction should be $\approx 90\%$ of the cosmic average for a galaxy-sized halo (\cite[Crain, Eke, Frenk \etal\ 2007]{Crain et al.}), on the other hand suggests the presence of some $\approx 1.4\times 10^{11}\ M_\odot$ of baryonic material within its virial radius, leaving around $60\%$ of it still to be found.

If some of the missing baryons of the Milky Way are locked up in the form of hydrogen-burning stars in a red halo, such a structure must somehow have evaded the faint star counts aimed to constrain the luminosity function of halo subdwarfs (e.g. \cite[Gould, Flynn \& Bahcall 1998]{Gould et al.}; \cite[Gould 2003]{Gould}; \cite[Digby \etal\ 2003]{Digby et al.}; \cite[Brandner 2005]{Brandner}), since no significant excess of low-mass stars have yet been detected with this method. We are currently investigating the red halo constraints imposed by observations of this kind (Zackrisson \& Flynn, in preparation). Preliminary results indicate that while a {\it uniform} halo population with an IMF as extreme as that envisioned by \cite[Zackrisson \etal\ (2006)]{Zackrisson et al.} can safely be ruled out, a halo with a strong radial population gradient (with an abnormal fraction of low-mass stars only at large Galactocentric distances) may be more difficult to disqualify. 

\section{Alternative explanations}
The idea of a halo population with an abnormally high fraction of low-mass stars is admittedly controversial, but so far, no alternative explanation has come close to explaining the red halo colours observed. The more mundane mechanisms that have been considered include: 
\begin{enumerate}
\item {\bf Nebular emission}. \cite[Zibetti \etal\ (2004)]{Zibetti et al.} suggest that the red halo colours could be caused by diffuse nebular emission (i.e. emission-lines and continuum associated with the ionized interstellar medium), and although such radiation may indeed be present at large distances from sites of active star formation, current models (\cite[Zackrisson \etal\ 2006]{Zackrisson et al.}) indicate that the spectral energy distribution of this component should be very blue, not red. This means that even if nebular emission is contributing to the red halo signal, it cannot be the origin of the extremely red colours observed. Correcting for this component would moreover render the underlying halo redder, and hence even more difficult to explain in the framework of a normal stellar population. Nebular emission may of course interfere with attempts to assess the extent and brightness of the red halo, and should preferably be corrected for. We are currently attempting to accomplish this by tracing the ionized component through narrowband surface photometry with filters centered on strong emission lines. 
\item {\bf High metallicity}. \cite[Rudy \etal\ (1997)]{Rudy et al.}, \cite{Bergvall & Östlin} and \cite[Zibetti \etal\ (2004)]{Zibetti et al.} have all suggested that the red halo colours may (at least partly) be explained by a stellar population with a high metallicity. However, as demonstrated by \cite[Zackrisson \etal\ (2006)]{Zackrisson et al.}, a high-metallicity population with a Salpeter-like IMF fails to explain the colours of the red halo detected around disk galaxies in stacked SDSS data. Due to the degeneracy between bottom-heavy IMFs and high metallicities in $BVK$ filters, a high-metallicity population with a normal IMF can in principle explain the red halos of BCGs, but this would require the metallicity to be \textit{very} high (Solar or higher), which is in conflict with the gaseous metallicities typically observed in BCGs ($\approx 10\%$ Solar). While the stellar population outside the bright star-forming centres of BCGs (where the gaseous metallicites are measured) may indeed be more metal-rich, the high metallicities inferred would require BCGs to be far more massive than indicated by current estimates (\cite[Bergvall \& \"Ostlin 2002]{Bergvall & Östlin}). This degeneracy between high stellar metallicities and bottom-heavy IMFs can be broken via $I$-band observations (\cite[Zackrisson \etal\ 2006]{Zackrisson et al.}), which have recently been obtained for a number of BCGs.
\item {\bf Dust reddening}. While dust reddening can, in principle, give rise to very red colours, normal dust reddening vectors (i.e. Milky Way, LMC, SMC dust) run almost parallel to the age vectors for old, normal-IMF stellar populations). While dust may be able to mimic a high age, it cannot explain the extreme colours observed in red halos. Assuming that the central star-forming regions of BCGs are encompassed from all angles by their red halos, the low extinction measured through the H$\alpha/$H$\beta$ emission-line ratios in BCGs (\cite[Bergvall \& \"Ostlin 2002]{Bergvall & Östlin}) moreover imposes a very strong upper limit on the amount of dust reddening in the halo. At least for these objects, dust reddening does not appear to be a viable explanation for the red halo phenomenon.
\item {\bf Dust emission}. Emission from hot dust could in principle help explain the $K$-band excess seen in the red halos of BCGs, but at least in the case of Haro 11 -- one of the BCGs with the reddest halo observed so far -- ISO observations have ruled out near-IR dust emission as an explanation for the colours observed (\cite[Bergvall \& \"Ostlin 2002]{Bergvall & Östlin}). Extended red emission from photoluminescent dust (\cite[e.g. Witt \& Vijh 2004]{Witt & Vijh}) could possibly contribute to the red excess of halos observed only in the optical (i.e. the red halo of stacked SDSS disks), but not likely to the red excess seen at longer wavelengths.  
\item {\bf Spectral synthesis problems}. The fact that a low-metallicity stellar population obeying a Salpeter-IMF fails to explain the observed red halos has been confirmed by several independent spectral evolutionary models. Recent updates related to the evolution of thermally pulsating asymptotic giant branch stars moreover seem to have but a modest impact on the interpretation of the colours used (\cite[Maraston 2005]{Maraston}; \cite[Bruzual 2007]{Bruzual}), and do not appear to be able to challenge this conclusion. The models moreover have no problem in explaining the colours of other red stellar populations like globular clusters or elliptical galaxies. There are, however, substantial uncertainties concerning the best-fitting age, metallicity and IMF slope derived from the red halo data in the framework of a bottom-heavy IMF, since current models tend to treat low-mass stars ($<0.8\ M_\odot$) in a very superficial manner or omit them completely. While justified in the case of a Salpeter-like IMF, such stars give a substantial contribution to the integrated light of bottom-heavy IMF populations, and should therefore be treated as carefully as possible. To alleviate this problem, a model of spectral evolution more suited for the prediction of the photometric properties of stellar populations obeying extreme IMFs is currently under development (Zackrisson, in preparation).
\item {\bf Sky subtraction problems}. The brightness of the night sky is about $\mu_B\approx 23 \ \mathrm{mag\ arcsec^{-2}}$ at the darkest sites on Earth, whereas red halos are detected at surface brightness levels of $\mu_B\approx 26$--27$\ \mathrm{mag\ arcsec^{-2}}$ for BCGs and $\mu_g\approx 28\ \mathrm{mag\ arcsec^{-2}}$ for the stacked SDSS disks. Hence, we are trying to measure brightnesses that may be as low as $1\%$ of the sky (and in the redder bands even less). The sky itself is also very red, implying that even a tiny undersubtraction of the sky could lead to a spurious detection of a very red structure. One way of testing for such sky subtraction problems is to insert synthetic images of galaxies (based on analytical surface brightness profiles, e.g. exponential or Sersic profiles, with known parameter values) into the observed sky frames and check whether the reduction routines used are able to return a surface brightness profile that is consistent with that used as input, out to the surface brightness levels where red halos are detected. Tests indicate that our sky subtraction software does indeed pass this trial. Bad things can and will happen once one attempts to approach surface brightness levels significantly below that of the extragalactic background light (Zackrisson \& \"Ostlin, in preparation), but since many of our red halo detections have been made at isophotes brighter than this, problems in properly correcting for the extragalactic background is unlikely to be the cause of the red halo phenomenon.
\item {\bf Point spread function effects}. One may also worry that the red halos may be the result of some instrumental artefact. Since red halos have so far only been detected in the outskirts of bright stellar systems, selectively scattered light from the central galaxies could in principle be a viable explanation for their unusual colours. \cite{Michard} argue that, depending on the quality of the telescope mirror, the point spread function (PSF) may indeed develop extended red wings which would give rise to diffuse, red structures in the vicinity of bright objects. While \cite[Zibetti \etal\ (2004)]{Zibetti et al.} estimate that the PSF may give non-negligible contribution to the halo signal observed in stacked SDSS data, they argue, based on measurements of SDSS PSF, that this cannot shift the colours substantially in the redward direction. Measurements of the PSF in the images where the red halos of BCGs have been detected have furthermore failed to reveal any similar red halos around stars, indicating that these detections are unlikely to result from PSF effects. Moreover, the red halos observed around BCGs sometime appear to be displaced relative to the BCG centre. To advocate PSF effects to explain such structures would require the PSF to be very asymmetric, and there is no evidence for that in the data. 
\end{enumerate}

\section{Red halo formation}
If the red halo phenomenon is indeed caused by a stellar population overly abundant in low-mass stars, how would such a structure have formed?  In the currently favoured scenario for stellar halo formation, the halo stars originally formed in less massive dark matter halos. As these low-mass halos fell into the potential well of a more massive galaxy, they disrupted due to tidal effects and redistributed their stellar content in the form of a diffuse halo of stars surrounding the centre of the more massive galaxy (see simulations by e.g. \cite[Bullock \& Johnston 2005]{Bullock \& Johnston}; \cite[Abadi, Navarro \& Steinmetz 2006]{Abadi et al.}). Simulations of this kind can of course only be expected to predict the gross features of such halos, since all the star formation details remain hidden on scales far below current resolution limits, and have to be inserted in the form of semi-analytical assumptions. While a red halo of low-mass stars is certainly not predicted by current simulations of this type, it is not obvious that such structures necessarily are inconsistent with this picture of stellar halo formation. It may indeed be possible to explain the existence of red halos in this framework, given suitable modifications of the star formation details that are inserted ``by hand''. 

At the current time, we can do little but speculate on the mechanisms responsible for the emergence of red halos. Here are a few scenarios that come to mind:
\begin{enumerate}
\item {\bf Population III stars}. Population III stars (i.e. stars with metallicity $Z\approx 0$) are expected to be distributed throughout the dark halo of a Milky Way-sized galaxy, albeit with a spatial distribution different from that of the cold dark matter (\cite[Scannapieco, Kawata, Brook, \etal\ 2006]{Scannapieco et al.}). Even though such stars are currently predicted to be very massive, any alternative paradigm for population III star formation that would instead preferentially generate low-mass stars could be a viable explanation for the formation of the red halos observed. 
\item {\bf A bottom-heavy IMF in dark galaxies}. Simulations based on $\Lambda$CDM generically predict that each galaxy-mass cold dark matter halo should contain a huge number of subhalos (typically accounting for $\approx 5$--10\% of its total mass) in the dwarf-galaxy mass range. Most of these subhalos do not, however, appear to correspond to luminous structures, as the Milky Way would then be surrounded by at least a factor of $\sim 10$ more satellite galaxies than observed, provided that each subhalo corresponds to a luminous dwarf galaxy (\cite[Klypin, Kravtsov, Valenzuela, \etal\ 1999]{Klypin et al.}; \cite[Moore, Ghigna, Governato, \etal\ 1999]{Moore et al.}). A similar lack of dwarf galaxies compared to the number of dark halos predicted is also seen on scales of galaxy groups (\cite[Tully, Somerville, Trentham, \etal\ 2002]{Tully et al.}). One way out of this dilemma is to assume that most of these low-mass halos correspond to so-called dark galaxies (e.g. \cite[Verde, Oh, \& Jimenez 2002]{Verde et al.}), i.e. objects of dwarf-galaxy mass which either do not contain baryons or in which the baryons formed very few stars. What if star formation does take place in these dark galaxies, but -- for whatever reason -- do so according to a bottom-heavy IMF? This would keep the dark galaxy population very dim indeed, and dispersion of their low-mass stars into the halo of a more massive galaxy could also help explain the red halos. This idea can be tested through very deep surface photometry of dark galaxy candidates, as these should then display very low surface brightness structures with a spectral energy distributions characteristic of a bottom-heavy IMF. In the ultra faint satellies recently discovered in the Milky Way, this may even be testable through direct star counts.
\item {\bf Dynamical mass segregation.} In principle, halo colours indicative of an abnormally high fraction of low-mass stars need not necessarily imply a bottom-heavy IMF. We may instead be witnessing the effects of dynamical mass segregation. Given sufficient time, equipartition of kinetic energy will tend to slow high-mass stars down and cause them to sink towards the centre of a stellar system, whereas low-mass stars will speed up and statistically end up at larger distances from the centre. Assuming that this process has had time to be efficient in each of the low-mass halos that eventually merged to form the high-mass systems that we witness today, one may envision that low-mass stars in the outer parts of these small halos were more easily stripped from these systems than the high-mass stars in the centre. This could give rise to a situation in which low-mass stars preferentially end up in the stellar halo, whereas high-mass stars tend to sink to the centre of the parent galaxy. While this could possibly explain the many low-mass stars that seem to inhabit the red halos, this scenario requires an excess of high-mass stars somewhere else in the galaxies that exhibit the red halo phenomenon.
\item {\bf In situ formation of low-mass stars}. A final option is of course that low-mass stars formed \textit{in situ} in the halos of these systems, possibly in a cooling flow (\cite[Mathews \& Brighenti 1999]{Mathews & Brighenti}).
\end{enumerate}
\section{Summary}
The red halos detected through deep optical/near-IR surface photometry of different types can currently only be understood by advocating a stellar halo population harbouring an abnormally high fraction of low-mass stars, possibly as an effect of a bottom-heavy IMF. Due to its high mass-to-light ratio, such a population effectively behaves as baryonic dark matter and could help explain the baryons still missing from current inventories relevant for the low-redshift Universe. Searches for red halos around types of galaxies have not yet been reported (post-starburst galaxies, elliptical galaxies and Local Group dwarfs) are currently underway, along with a number of observational tests aimed to constrain the nature of the red halos. 

\begin{acknowledgments}
EZ acknowledges research grants from the Swedish Research Council, the Academy of Finland and the Swedish Royal Academy of Sciences. CF acknowledges support from the Academy of Finland.
\end{acknowledgments}

\end{document}